\documentclass[prd,aps,tightenlines,twocolumn,nofootinbib,preprintnumbers]{revtex4}
\usepackage{amsmath,amssymb}
\usepackage{epsfig}
\usepackage{graphicx}
\usepackage{subfigure}

\usepackage{slashed}

\newcommand{\be}{\begin{equation}}
\newcommand{\ee}{\end{equation}}
\newcommand{\bea}{\begin{eqnarray}}
\newcommand{\eea}{\end{eqnarray}}
\newcommand{\mg}{m_{3/2}}
\newcommand{\g}{\widetilde{G}}

\newcommand{\mpl}{M_{Pl}}
\newcommand{\half}{\frac{1}{2}}

\newcommand{\lsim}{\mbox{\raisebox{-.6ex}{~$\stackrel{<}{\sim}$~}}}
\newcommand{\gsim}{\mbox{\raisebox{-.6ex}{~$\stackrel{>}{\sim}$~}}}
\newcommand{\G}{\mathrm{GeV}}

\begin{document}

\title{Non-perturbative production of matter and rapid thermalization after MSSM inflation}
\author{Rouzbeh Allahverdi~$^1$, Andrea Ferrantelli~$^{2,3}$, Juan Garcia-Bellido~$^{4,5}$, Anupam Mazumdar~$^{6,7}$}

\affiliation{$^1$~Department of Physics and Astronomy, University of New Mexico, Albuquerque, NM 87131, USA\\
$^2$~Helsinki Institute of Physics, P.O. Box 64, FI-00014 University of Helsinki, Finland
\\
$^3$~Department of Structural Engineering and Building Technology, Aalto University, FI-02150 Espoo, Finland
\\
$^4$~Instituto de F\'{i}sica Te\'{o}rica UAM-CSIC, Universidad Aut\'{o}noma de Madrid, Cantoblanco, 28049 Madrid, Spain\\
$^5$D\'epartement de Physique Th\'eorique, Universit\'e de Gen\`eve, CH-1211 Gen\`eve 4, Switzerland\\
$^6$~Physics Department, Lancaster University, Lancaster, LA1 4YB, United Kingdom\\
$^7$~Niels Bohr Institute, Copenhagen University, Blegdamsvej-17, DK-2100, Denmark}


\begin{abstract}
A {\it gauge invariant} combination of LLe {\it sleptons} within the Minimal Supersymmetric Standard Model is one of the few inflaton candidates that can naturally explain
population of the observable sector and creation of matter after inflation.
After the end of inflation, the inflaton oscillates coherently about the minimum of its potential,
which is a point of {\it enhanced gauged symmetry}.
This results in bursts of non-perturbative production of the gauge/gaugino and (s)lepton quanta. The subsequent decay of these quanta is very fast and leads to an extremely efficient transfer of the inflaton energy to (s)quarks via {\it instant} preheating. Around $20\%$ of the inflaton energy density is drained during every inflaton oscillation. However, all of the Standard Model degrees of freedom (and their supersymmetric partners)  {\it do not} thermalize immediately, since the large inflaton vacuum expectation value breaks the electroweak symmetry.
After about 100 oscillations -- albeit within one Hubble time -- the amplitude of inflaton oscillations becomes sufficiently small, and all of the degrees of freedom will thermalize. This provides by far the most efficient reheating of the universe with the observed degrees of freedom.
\end{abstract}

\preprint{IFT-UAM/CSIC-11-08}
\preprint{HIP-2011-06/TH \hspace{7.5mm}}

\maketitle


\section{Introduction}

Primordial inflation~\cite{Guth} has many virtues -- it explains the large scale homogeneity and stretches the initial seed perturbations
to the observable scales in the cosmic microwave background radiation~\cite{WMAP}. In spite of the observational successes, a prime question remains regarding the microphysical origin of inflation -- what is the inflaton?

There are many inflaton candidates, see e.g.~\cite{RM}, which are fully capable of explaining the cosmic microwave background radiation, but it is not clear how they would create the {\it observed} fundamental particles from the inflaton-induced vacuum energy. Reheating after inflation must generate all the Standard Model (SM) degrees of freedom required for the success of Big Bang Nucleosynthesis~\cite{BBN} (for a review on reheating, see e.g.~\cite{ABCM}.).

Inflation can occur within a hidden sector, where the inflaton is a Standard Model (SM) gauge singlet, and couples to all of the hidden and visible sector degrees of freedom. However, the process of transferring the energy from the inflaton and all the hidden sectors to the visible sector degrees of freedom is largely unknown due to the fact that the couplings are mostly unknown beyond the strong, weak and electromagnetic interactions. Furthermore, whether all the hidden degrees of freedom can decay into the visible sector or not is a debatable issue.\footnote{In order to answer all these questions, one must embed the hidden sector inflation within a known setup where both UV and IR aspects of the theory are well understood. Only one such example is known throughout the literature~\cite{Michele}, where a closed string modulus within Large Volume Compactifications is responsible for generating the seed perturbations, and {\it all} possible couplings of the inflaton to hidden and visible sectors are known. In this setup, the majority of the inflaton energy density is driven to excite the hidden sector particles as compared to the matter fields. One requires stringent and very speculative constraints to overcome the problem~\cite{Michele}.}

These issues can be resolved if the inflaton belongs to the visible sector, and its mass and couplings to the SM fields are well known.
It has been demonstrated~\cite{AEGM,AKM,AEGJM} that inflation can occur within the Minimal Supersymmetric Standard Model (MSSM) and its minimal extensions, with the remarkable property that the inflaton is {\it not} an arbitrary gauge singlet. Rather, it is a $D$-flat direction in the scalar potential consisting of the supersymmetric partners of quarks and leptons.\footnote{For a review on MSSM flat directions, see~\cite{MSSM-REV,Dine}}. These models give rise to a wide range of scalar spectral indices~\cite{AEGJM,BDL,EMS,HMN}, including the whole range allowed by WMAP~\cite{WMAP}~\footnote{ Although it was pointed out in original papers that there exists tuning of
the MSSM parameters at the scale of inflation to maintain the flatness of
the potential~\cite{AEGM,AKM,AEGJM}. However in subsequent studies it was noted that this tuning
can be ameliorated substantially not only at high scales, see Refs.~\cite{EMS,HMN,Mazumdar:2011ih}, but
also at scales where squark and selpton masses would be measured at the
LHC, see Ref.~\cite{ADS}. The initial condition problem has been addressed in Refs.~\cite{AFM,ADM3}.}.

Moreover, since the inflaton is an MSSM flat direction, its couplings to matter fields are known. It is therefore possible to track the thermal history of the universe from the end of inflation.\footnote{The parameter space permitting successful inflation is compatible with supersymmetric dark matter~\cite{ADS,ADM2} (and may even lead to a unified origin of inflation and dark matter~\cite{ADM1}).} After the end of inflation, the flat direction starts coherent oscillations around its minimum, denoted by $\phi=0$, which is also a point of {\it enhanced gauge symmetry}~\cite{AEGJM}.  
Whenever the inflaton passes through the origin, the eigenmodes of the fields that are directly coupled to the inflaton are excited via a non-perturbative phenomenon~\cite{TB,KLS,STB,KLS1}.
At values of the inflaton away from the minimum, the same modes become heavy and therefore it is kinematically unfavorable to excite them. However, they can and will decay to particles that are coupled to them (but not to the inflaton).
This is the process known as {\it instant preheating}~\cite{instant}.

This scenario has been briefly discussed in its generality in~\cite{AEGJM}. In this paper, we present a detailed investigation for the case that the $LLe$ flat direction (which consists of sleptons) plays the role of the inflaton. In this case the inflaton passage through the origin results in non-perturbative production of the electroweak gauge/gaugino and (s)lepton quanta. These will subsequently decay to (s)quarks very quickly. As we will show, draining the inflaton energy via instant preheating is quite efficient, and nearly $10\%$ of the inflaton energy density gets transferred to the relativistic particles at every zero-crossing~\footnote{To be clear, our discussions concerns creation of particles and antiparticles associated with matter fields during reheating. Generation of matter-antimatter asymmetry happens at a later stage.}.

However, since the inflaton Vacuum Expectation Value (VEV) breaks the electroweak symmetry, all of the MSSM degrees of freedom will not reach thermal equilibrium immediately. After about 100 oscillations, the amplitude of inflaton oscillations becomes sufficiently small and electroweak interactions become efficient. At this point full equilibrium is achieved. Due to the hierarchy between the frequency of oscillations $\sim {\cal O}({\rm TeV})$ and the expansion rate $\sim {\cal O}({\rm GeV})$, the universe thermalizes within one Hubble time after the end of inflation.


This paper is structured as follows. We begin by presenting a general discussion of MSSM inflation focusing on the LLe inflaton in section 2.
In section 3, we derive the inflaton couplings to scalar, gauge and fermion fields. We discuss reheating after MSSM inflation and the dominance of instant preheating in transferring the energy away from the inflaton oscillations in section 4. In section 5, we discuss the subsequent stage of thermalization of the MSSM degrees of freedom, and give an estimate of the reheat temperature. Finally, we conclude our paper in section 6. Essential expressions for the $LLe$ inflaton and the relevant interactions for the scalar, gauge boson and fermion fields are given in the Appendix.


\section{The ${\bf LLe}$ inflaton}

We concentrate on the ${LLe}$ flat direction as an inflaton.\footnote{The other possible candidates are ${\bf udd}$ (where
${u,~d}$ correspond to right-handed squarks~\cite{AEGM,AEGJM}), and ${NH_{u}L}$ (where ${\bf N}$ corresponds to the right-handed sneutrino and ${H_{u}}$ is the MSSM Higgs doublet that gives mass to up-type quarks~\cite{AKM}). We will not discuss these cases in this paper.} Here ${L}$ and ${e}$ denote the left-handed and right-handed lepton superfields, which are doublet and singlet under the $SU(2)_W$, respectively, and carry hypercharge quantum numbers $-1$ and $+2$, respectively, under the $U(1)_Y$. The $D$- and $F$-flatness conditions require that the family indices of the three superfields be different. Without loss of generality, we can choose the indices as $L_1 L_2 e_3$. Then the flat direction background can be parameterized as
\begin{equation} \label{flat}
{\tilde L}_1 = {1 \over \sqrt{3}}
\left(\begin{array}{ll}
\varphi \\ 0 \end{array}\right), ~~
{\tilde L}_2 = {1 \over \sqrt{3}}
\left(\begin{array}{ll}
0 \\ \varphi \end{array}\right), ~~ {\tilde e}_3 = {1 \over \sqrt{3}} \varphi.
\end{equation}
Here ${\tilde L},~{\tilde e}$ are scalar components of the corresponding superfields and $\varphi$ is a complex scalar fields.
The inflaton field $\phi$ will be identified by the real part of $\varphi$: $\phi = \varphi_R$.

The scalar potential along $\phi$ is given by~\cite{AEGM,AEGJM}:\footnote{For a detailed dynamics of inflection point inflation, see~\cite{EMS}.}
\begin{equation} \label{flatpot}
V(\phi) = {1 \over 2} m^2_\phi \phi^2 - A \lambda {\phi^6 \over 3 M^{3}_{\rm P}} + \lambda^2 {{\phi}^{10} \over M^{6}_{\rm P}} \, .
\end{equation}
For $A \approx \sqrt{40} m_{\phi}$ there exists an inflection point at
\begin{equation} \label{saddle}
\phi_0 = \left({m_\phi M^{3}_{\rm P} \over \sqrt{10} \lambda}\right)^{1/4} ,
\end{equation}
in the potential. Successful inflation can occur within an interval $\vert \phi - \phi_0 \vert \sim (\phi^3_0 / 60 M^2_{\rm P})$ in the vicinity of $\phi_0$.

For $m_{\phi} \sim {\cal O}({\rm TeV})$, as preferred by weak scale supersymmetry, we have $\phi_0 \sim {\cal O}(10^{14})$ GeV, and the Hubble expansion rate during inflation is $H_{\rm inf} \sim {\cal O}({\rm GeV})$.

After inflation the inflaton starts oscillating about $\phi = 0$ with a frequency $m_\phi$.
The inflaton potential, see Eq.~(\ref{flatpot}), has two points of inflection,\footnote{Because of the potential is symmetric under $\phi \rightarrow -\phi$, there are two other points of inflection at $-\phi_0$ and $-\phi_1$.}
%
\be \label{1stinf}
\phi_0 ~ , ~ \phi_1 = {\phi_0 \over \sqrt{3}} ,
\ee
%
with the respective potential energies
\be \label{2ndinf}
V(\phi_0) = {4 \over 15} m^2_\phi \phi^2_0 ~ ~ , ~ ~ V(\phi_1) = {V(\phi_0) \over 10} .
\ee
The potential is tachyonic $-2 m^2_\phi \leq V^{\prime \prime} (\phi) \leq 0$ between the two inflection points, while for $\phi < \phi_1$ we have $V^{\prime \prime} > 0$.

An important point to keep in mind is that because of the hierarchy $m_\phi \sim 10^3 H_{\rm inf}$, the inflaton can undergo a large number of oscillations within a single Hubble time after the end of inflation. As we will see later, this has interesting consequences for reheating in this model.

\begin{figure}[htb]
\begin{center}
\begin{tabular}{c}
\includegraphics[width=8.5cm]{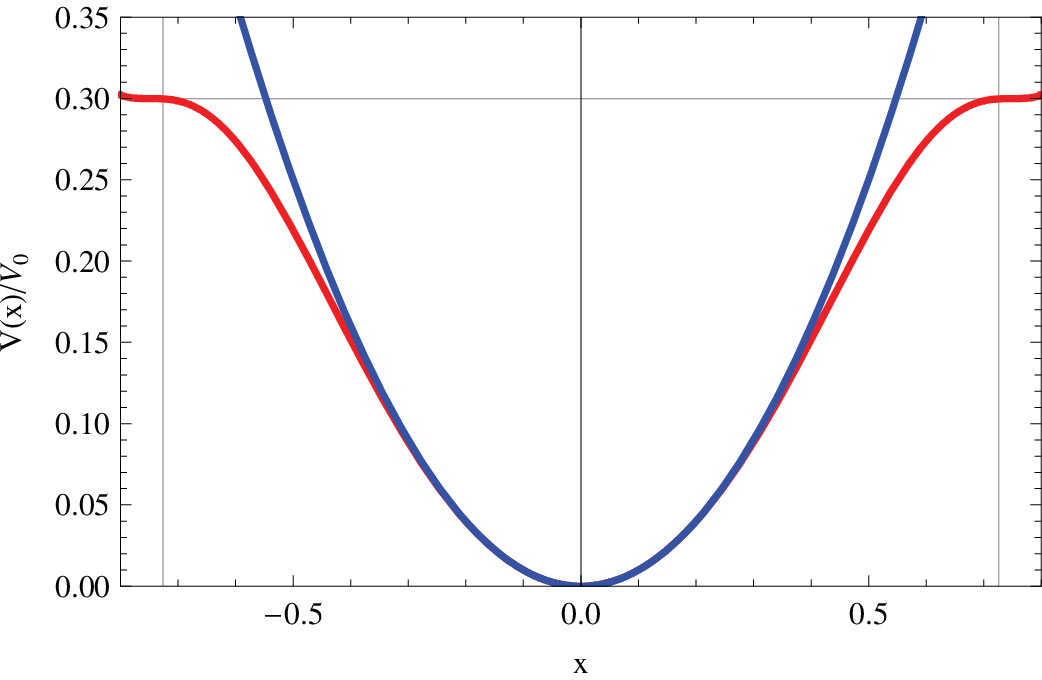}
\end{tabular}
\caption{The inflaton potential around the minimum, together with the first approximation, $m_\phi^2\phi^2$. The variable $x=10^{1/8}\phi/\phi_0$.}
\label{potential}
\end{center}
\end{figure}


\section{Couplings of the inflaton}

Any time that the inflaton crosses the origin scalar, gauge and fermion fields that are coupled to it are produced non-perturbatively. Production of particles that have couplings of gauge strength to the inflaton dominates over that of particles with Yukawa couplings to the inflaton. After the inflaton passes the origin it rolls back to large VEVs, and the produced particles become very heavy. They can therefore decay quickly to particles that are lighter than them (i.e. particles with no gauge couplings to the inflaton). For the $L_1 L_2 e_3$ inflaton, these include all of the (s)quarks, the Higgs and Higgsino particles, and $L_3,~e_1,~e_2$ (s)leptons. These light fields interact among themselves, which may lead to their thermalization.
For a proper treatment of these processes, we first need to identify different fields and their couplings to each other.


\subsection{Coupling to scalars and their decay widths}

We focus on the scalars that have gauge couplings to the inflaton as they play the dominant role in transferring the inflaton energy away. The relevant interaction terms arise from the $D$-term part of the scalar potential.
The field content of the ${\tilde L}_1, {\tilde L}_2, {\tilde e}_3$ is
\begin{equation}\label{scal}
{\tilde L}_1 = \left(\begin{array}{ll}
\varphi_1 \\ \varphi_2 \end{array}\right), ~~
{\tilde L}_2 = \left(\begin{array}{ll} \varphi_3 \\ \varphi_4 \end{array}\right),
~~ {\tilde e}_3 = \varphi_5 ,
\end{equation}
which includes 10 real degrees of freedom in total. For the background given in Eq.~(\ref{flat}), the inflaton $\phi$ is
\begin{equation} \label{inflaton}
\phi = {\varphi_{1,R} + \varphi_{4,R} + \varphi_{5,R} \over \sqrt{3}}.
\end{equation}
The combination $(\varphi_{1,I} + \varphi_{4,I} + \varphi_{5,I})/\sqrt{3}$ has a mass $m_\phi$ and is irrelevant for inflation ($R,~I$ denote the real and imaginary parts of a complex scalar field respectively).

The inflaton VEV completely breaks the $SU(2)_W \times U(1)_Y$ symmetry. This results in four massive real scalars
\begin{eqnarray} \label{scaldef}
&& \chi_1 = {\varphi_{2,R} + \varphi_{3,R} \over \sqrt{2}} \, , ~ \chi_2 =  {\varphi_{2,I} - \varphi_{3,I} \over \sqrt{2}} \, , ~ \chi_3 = {\varphi_{1,R} - \varphi_{4,R} \over \sqrt{2}} \, , \nonumber \\
&& \chi_4 = \sqrt{2 \over 3} ({\varphi_{5,R} - {1 \over 2} \varphi_{1,R} - {1 \over 2} \varphi_{4,R}}) \, ,
\end{eqnarray}
whose masses are obtained from the $D$-terms (for detailed derivation, see Appendix~\ref{DWOS}). From Eqs.~(\ref{scalint1},\ref{scalint2}) we find the following mass terms
\begin{eqnarray}\label{scalcoupl}
V  \supset  {1 \over 12} g_W^2 \phi^2 (\chi^2_1 + \chi^2_2  +  \chi^2_3) + {1 \over 4} g_Y^{ 2} \phi^2 \chi^2_4 \, .
\end{eqnarray}
Here $g_W,~g_Y$ are the $SU(2)_W$ and $U(1)_Y$ gauge couplings respectively.

The remaining four degrees of freedom that are orthogonal to $\phi$ (and its imaginary counterpart) and $\chi$'s are Goldstone bosons from breakdown of $SU(2)_W \times U(1)_Y$. They are eaten by the Higgs mechanism and give rise to longitudinal components of the electroweak gauge fields. In the unitary gauge, they are completely removed from the spectrum.

The $\chi$ particles decay to squarks, the Higgs particles, and the ${\tilde L}_3,~{\tilde e}_1,~{\tilde e}_2$ sleptons.
The relevant interaction terms between $\chi$'s and these fields are given in Appendix~\ref{DWOS}.
From Eq.~(\ref{chidec}), and after summing over all decay channels, we find
\begin{eqnarray} \label{scaldec}
\Gamma_{\chi_1} = \Gamma_{\chi_2} = \Gamma_{\chi_3} = {3 g_W^3 \phi \over 8 \pi \sqrt{6}} ~ , ~
\Gamma_{\chi_4} = {9 g_Y^{3} \phi \over 16 \pi \sqrt{2}} \, .
\end{eqnarray}
Note that the decay rate is proportional to the VEV of the inflaton, which sets the mass of $\chi$ fields.


\subsection{Coupling to gauge fields and their decay widths}

Couplings of the inflaton to the gauge fields are obtained from the flat direction kinetic terms (see Appendix \ref{DWOGF}). From Eqs.~(\ref{kin},\ref{der}), and after using Eq.~(\ref{inflaton}), we find the following mass terms for the electroweak gauge fields
\begin{eqnarray} \label{gaugcoupl}
{\cal L} \supset {1 \over 12} g_W^2 \phi^2 (2 W^{+, \mu} W^{-}_{\mu} + W^{\mu}_3 W_{3,\mu}) + {1 \over 4} g_Y^{ 2} \phi^2 B^{\mu} B_{\mu} \, , \nonumber \\
\,
\end{eqnarray}
where
\begin{equation} \label{w}
W^{+} = {W_1 - i W_2 \over \sqrt{2}} ~ ~ ~ , ~ ~ ~ W^{-} = {W_1 + i W_2 \over \sqrt{2}},
\end{equation}
and $W_{i,\mu}$ and $B_\mu$ are the $SU(2)_W$ and $U(1)_Y$ gauge fields respectively.
The gauge fields decay to (s)quarks, Higgs and Higgsino particles, and $L_3,~e_1,~e_2$ (s)leptons.
The relevant interaction terms are given in Eqs.~(\ref{gaugscint},
\ref{gaugfeint}). The total decay widths of gauge fields are found to be
\begin{eqnarray} \label{gaugdec}
\Gamma_{W^+} = \Gamma_{W^-} = \Gamma_{W_3} = {3 g_W^3 \phi\over 8 \pi \sqrt{6}} ~ , ~
\Gamma_B = {9 g_Y^{ 3} \phi\over 16 \pi \sqrt{2}}  \, ,
\end{eqnarray}
where we have used Eq.~(\ref{gaugcoupl}) for the gauge field masses.


\subsection{Coupling to fermions and their decay widths}

Couplings of the inflaton to fermions are found Eq.~(\ref{fermlag}), see Appendix~\ref{DWOF}. They result in the following mass terms for fermions
\begin{eqnarray} \label{fermcoupl}
{\cal L} \supset {1 \over \sqrt{6}} g_W \phi ({\bar \Psi}_1 {\Psi}_1 +  {\bar \Psi}_2 {\Psi}_2 + {\bar \Psi}_3 {\Psi}_3) + {1 \over \sqrt{2}} g_Y \phi {\bar \Psi}_4 {\Psi}_4 \, , \nonumber \\
\,
\end {eqnarray}
where $\Psi_1,~\Psi_2,~\Psi_3,~\Psi_4$ are Dirac spinors defined in Eq.~(\ref{fermdef}).
The $\Psi$'s decay to (s)quarks, Higgs and Higgsaino particles, and $L_3,~e_1,~e_2$ (s)leptons.
The relevant interaction terms are given in Eq.~(\ref{fermint}). After summing over all final states, we find the following decay widths
\begin{eqnarray} \label{fermdec}
\Gamma_{\Psi_1} = \Gamma_{\Psi_2} = \Gamma_{\Psi_3} = {3 g_W^3 \phi \over 8 \pi \sqrt{6}} ~ , ~
\Gamma_{\Psi_4} = {9 g_Y^{3} \phi \over 16 \pi \sqrt{2}}\,,
\end{eqnarray}
where we have used Eq.~(\ref{fermcoupl}) for the mass of $\Psi$'s.

Note that $\chi_{1,2,3}$ scalars, $SU(2)_W$ gauge fields $W^{\pm},~W_3$, and $\Psi_{1,2,3}$ fermions have the same mass, see Eqs.~(\ref{scalcoupl},\ref{gaugcoupl},\ref{fermcoupl}), and the same decay width, see Eqs.~(\ref{scaldec},\ref{gaugdec},\ref{fermdec}). Similarly, $\chi_4$ scalar, $U(1)_Y$ gauge field $B$, and $\Psi_4$ fermion have the same mass and decay width. This is expected from supersymmetry. The inflaton VEV results in supersymmetry conserving masses for the scalars, gauge fields and fermions. After the breakdown of $SU(2)_W \times U(1)_Y$, the bosonic degrees of freedom include four real scalars $\chi_i$ and four massive gauge fields, and their fermionic partners are grouped into four Dirac spinors.


\section{Reheating after MSSM inflation}

We now discuss the reheating stage after MSSM inflation and various channels for energy transfer from the inflaton to SM quarks and leptons and their supersymmetric partners. We will focus on the various ways of exciting the light and heavy degrees of freedom.

After inflation, the inflaton starts oscillating about the origin with an initial amplitude ${\hat \phi}_0 \simeq \phi_0$. There are various mechanisms for particle production from an oscillating condensate. Here we discuss those that are relevant for the MSSM inflaton.


\subsection{Tachyonic preheating}

As pointed out before, the inflaton potential is tachyonic $V^{\prime \prime}(\phi) < 0$ for $(\phi_0/\sqrt{3}) < \phi < \phi_0$. Quantum fluctuations of the non-zero inflaton modes are amplified due to tachyonic instabilities as the field sweeps this interval~\cite{tachy}. This effect will go away once the amplitude of oscillations ${\hat \phi}$ drops below $\phi_0/\sqrt{3}$, which amounts to transfer of $10\%$ of the energy density in the oscillating condensate. A question arises that to what extent tachyonic preheating is efficient in reducing the energy of the inflaton zero mode by this amount.

Examination of the potential~(\ref{flatpot}) shows that $-2m^2_\phi \leq V^{\prime \prime}(\phi) \leq 0$ within the $[(\phi_0/\sqrt{3}),\phi_0]$ interval. Therefore modes with momentum $0 \leq k^2 \leq 2 m^2_\phi$ can undergo amplification because of tachyonic instabilities,
i.e. $n_{k}\sim e^{4m_{\phi} t_{\ast}}$, where $t_{\ast}$ is the time spent during the tachyonic phase~\cite{tachy}. The amount of amplification needed to increase the energy density in these modes from its initial value ${\cal O}(m^4_\phi)$ to $10^{-1} V(\phi_0) \sim 10^{-2} m^2 \phi^2_0$ is ${\cal O}(10^{12})~{\rm GeV^4}$. This would require at least ${\cal O}(10)$ oscillations.

On the other hand, as shown below, instant preheating transfers ${\cal O}(20\%)$ of the inflaton energy to relativistic particles in just one oscillation, which is mainly due to the hierarchically larger phase space for resonant particle production, see Eq.~(\ref{momentum}). This implies that instant preheating reduces ${\hat \phi}$ below $\phi_0/\sqrt{3}$ after ${\cal O}(1)$ oscillations, after which tachyonic preheating will be irrelevant since $V^{\prime \prime} > 0$ at all times, see Figs.~1 and 2.

We therefore conclude that tachyonic preheating ends before it can become competitive with instant preheating in draining the energy density from the inflaton zero mode.


\subsection{Instant preheating}

The fields that are coupled to the inflaton acquire a VEV-dependent mass that varies in time due to the inflaton oscillations. For illustration, we first focus on the $\chi_1$ scalar, see Eq.~(\ref{scalcoupl}).\footnote{Similar arguments and calculations will hold for gauge bosons and fermions. Single-crossing occupation numbers in instant preheating are insensitive to the spin of the field.} The $\chi_1$ quanta are produced every time the inflaton passes through the origin~\cite{instant}. The Fourier eigenmodes of $\chi_1$ have the corresponding energy
\begin{eqnarray} \label{eigenmodes}
\omega_k &\!=\!& \sqrt{k^2 + m_{\chi_1}^2 + {g^2_W {\phi (t)}^2/6}} \nonumber \\
&\!=\!& \sqrt{k^2 + m_{\chi_1}^2 + 4 m_\phi^2 \tau^2 q_1}\,,
\end{eqnarray}
with $\phi(t)$ being the instantaneous VEV of the inflaton and $m_{\chi_1}$ the bare (time-independent) mass of the $\chi_1$ field, and we have written it in terms of the broad resonance parameter
\begin{equation}\label{q}
q_1 \equiv\frac{g^2_W \dot\phi_0^2}{24m_\phi^4}\gg 1\,,
\end{equation}
and the time $\tau=m_\phi t$ after the inflaton zero-crossing.
As is well known, a given mode gets excited when the adiabaticity condition is violated ${\dot \omega}_{k} \gsim \omega^2_{k}$. This happens for modes with $k \lsim k_{\rm max}$ each time the inflaton crosses the origin, where
\be \label{momentum}
k^2_{\rm max} \simeq {1 \over\sqrt6} g_W \dot \phi_0 = 2m_\phi^2\,\sqrt q_1\,.
\ee
Here ${\dot \phi}_0$ is the velocity of the inflaton at zero crossing.\footnote{Neglecting the expansion of the universe, we have ${\dot \phi}_0 = \sqrt{2 V({\hat \phi})}$ from the conservation of energy, where ${\hat \phi}$ is the amplitude of the inflaton oscillations. Note that after a few oscillations, $\dot\phi_0 \simeq m_\phi\hat\phi$, see Fig.~2.} The growth of the occupation number of mode $k$ can be computed exactly for the first zero-crossing,
\bea \nonumber
n_{k,\chi_1} &\!=\!& \exp\!\left[{-\frac{\pi\sqrt6(k^2+m_{\chi_1}^2)}{g_W \dot\phi_0}}\!\right] \\ \label{nk}
&\!=\!& \exp\!\left[{-\frac{\pi(k^2+m_{\chi_1}^2)}{2m_\phi^2\sqrt q_1}}\!\right] < 1\,,
\eea
The total number density of particles thus produced follows
\bea \nonumber
n_{\chi_1} \!&=&\! \int_{0}^\infty\frac{d^3 k}{(2 \pi)^3} \exp\left[-\frac{\pi(k^2+m_{\chi_1}^2)}{2m_\phi^2\sqrt q_1}\right] \\
\!&=&\!  {m_\phi^3 \over 2\sqrt2 \pi^3} q^{3/2}_1\,\exp\left({-\frac{\pi m_{\chi_1}^2}{2m_\phi^2\sqrt q_1}}\right) \,. \label{ndensity}
\eea
This expression corresponds to the asymptotic value and assumes there is no perturbative decay of the produced $\chi$ particles. However, immediately after adiabaticity is restored,
\be
t>t_{*,1}=\sqrt{\frac{\sqrt6}{g_W\dot\phi_0}} \hspace{2mm}\Rightarrow\hspace{2mm} \tau_1 > \tau_{*,1} = \frac{1}{\sqrt2}\,q^{-1/4}_1\, ,
\ee
$\chi_1$ particles can and will decay into lighter particles (i.e. those particles that have no gauge coupling to the inflaton). As mentioned earlier, in the case of $L_1 L_2 e_3$ inflaton these are the (s)quarks, Higgs(inos), and $L_3,~e_1,~e_2$ (s)leptons. The transfer of the inflaton energy to these light particles is called instant preheating~\cite{instant}.

As the inflaton is rolling back the potential, the instantaneous mass that it induces for the produced $\chi_1$ quanta increases, and so does their decay width $\Gamma_{\chi_1}$.
At the time of decay we have $\phi(t_{\rm dec,1}) \simeq \dot \phi_0\, t_{\rm dec,1}$ (assuming $t_{\rm dec,1} \ll m^{-1}_{\phi}$), where $t_{\rm dec,1} \sim \Gamma^{-1}_{\chi_1}$ denotes the time between the zero crossing and $\chi_1$ decay. After using Eq.~(\ref{scaldec}), we find
\be \label{dectime}
1 \gg \tau_{\rm dec,1} = m_\phi t_{\rm dec,1} \sim \left(\frac{4\pi}{3g_W^2}\right)^{1/2}\,q^{-1/4}_1 \gg \tau_{*,1} \, .
\ee
For typical values of $m_\phi \sim 100-1000$ GeV, we have ${\hat \phi} \sim \phi_0 \simeq 10^{14}-10^{15}$ GeV, at the beginning of oscillations, which results in $t_{\rm dec,1} \sim 6 \times 10^{-6} m^{-1}_\phi$. This confirms the validity of our approximation and ensures that $\chi_1$'s indeed decay promptly after their production. We also note that $g_W \phi^2(t_{\rm dec,1}) \gg {\dot \phi}_0$, which ensures that $\chi_1$'s are non-relativistic at the time of decay, and hence using the decay rate in the $\chi_1$ rest frame~(\ref{scaldec}) is valid.

The energy density in $\chi_1$ particles soon after zero crossing is given by
\bea \nonumber
\rho_{\chi_1}(\tau) &\!=\!& \int_0^\infty \frac{k^2\,dk}{2\pi^2} n_{k,\chi_1}\,\omega_k(\tau)\\
&\!=\!&  \frac{q_1\,m_\phi^4}{\pi^4}\,A_1 e^{A_1} K_1(A_1)\,\exp\left({-\frac{\pi m_{\chi_1}^2}{2m_\phi^2\sqrt q_1}}\right) \, , \nonumber \\
\, \label{edensity}
\eea
where
\begin{eqnarray}
A_1 \equiv \frac{\pi m_{\chi_1}^2}{4 \sqrt q_1 m_\phi^2}+ \pi \sqrt q_1 \, \tau^2 \simeq \pi\sqrt q_1 \, \tau^2 \, ,
\end{eqnarray}
and $K_1(z)$ is the modified Bessel function of the second kind, satisfying $zK_1(z) = 1 +{\cal O}(z^2)$.

$\rho_{\chi_1}$ actually decays with the decay rate $\Gamma_{\chi_1}$,
\bea \nonumber
\rho_{\chi_1}(\tau) &\!=\!& \rho_{\chi_1}\,{\rm exp}\!\left[{-\int_0^\tau \Gamma_{\chi_1} dt}\!\right] \\
&\!=\!&
\frac{q_1\,m_\phi^4}{\pi^4}\,A_1 e^{A_1} K_1(A_1)\, e^{-A_1/2A_{\rm dec,1}} \,
\exp\!\left[{-\frac{\pi m_{\chi_1}^2}{2m_\phi^2\sqrt q_1}}\!\right] \, \nonumber \\
&& \,
\eea
with $A_{\rm dec, 1} \equiv A_1(\tau_{\rm dec})= (4\pi^2/3g_W^2) \simeq 36.6$, which can be integrated beyond the time of decay, to give
\bea \label{densitydec}
\bar\rho_{\chi_1} \!\simeq\! 7.82 \times \frac{q_1\,m_\phi^4}{\pi^4}\,\exp\!\left[{-\frac{\pi m_{\chi_1}^2}{2m_\phi^2\sqrt q_1}}\!\right] \, ,
\eea
with $\rho_\phi=\dot\phi_0^2/2$ the inflaton energy at each zero-crossing. Thus the fraction that is transferred from the inflaton to $\chi_1$'s, and through their prompt decay into relativistic squarks at every inflaton zero-crossing, is given by 
\be \label{transfer}
{\bar\rho_{\chi_1} \over \rho_\phi} \sim 0.0067\,g_W^2\,\exp\!\left[{-\frac{\pi m_{\chi_1}^2}{2m_\phi^2\sqrt q_1}}\!\right]\,.
\ee
The expressions as in Eqs.~(\ref{dectime},\ref{densitydec},\ref{transfer}) also hold for $\chi_2,~\chi_3$ scalars, $W^+,W^-,W_3$ gauge bosons, and $\Psi_1,~\Psi_2,~\Psi_3$ fermions. The reason being that all of these fields have the same mass and decay width as $\chi_1$, and single-crossing occupation numbers in instant preheating are insensitive to the spin of the field.\footnote{The fraction of the inflaton energy transferred via Yukawa couplings is given by Eq.~(\ref{transfer}), with the gauge coupling $g_W$ replaced by a Yukawa coupling. It is seen from the parametric form of this expression that the fraction becomes smaller for smaller couplings. This is why we can safely ignore all fields that have only Yukawa couplings to the inflaton.}

On the other hand, the $\chi_4$ scalar, $B$ gauge field, and $\Psi_4$ fermion have a different mass and decay rate. The corresponding expressions for them are
\be
q_4 \equiv \frac{g_Y^2\dot\phi_0}{8m_\phi^2}\, ,
\ee
and
\be \label{dectime4}
1 \gg \tau_{\rm dec,4} = m_\phi t_{\rm dec,4} \sim \left(\frac{8\pi}{9g_Y^2}\right)^{1/2}\,q_4^{-1/4} \gg \tau_{*,4}
\ee
which gives the same expression for the number density and energy density in $\chi_4$ particles, but with $q=q_4$ and $A_{\rm dec,4}=(8\pi^2/9g_Y^2) \simeq 24.4$,
\be
n_{\chi_4} = {m_\phi^3 \over 2\sqrt2 \pi^3} q^{3/2}_4\,\exp\left({-\frac{\pi m_{\chi_4}^2}{2m_\phi^2\sqrt q_4}}\right) \, , \label{ndensity4}
\ee
and
\be
\bar\rho_{\chi_4} \simeq 6.46 \times \frac{q_4\,m_\phi^4}{\pi^4}\,\exp\!\left[{-\frac{\pi m_{\chi_4}^2}{2m_\phi^2\sqrt q_4}}\!\right] ,
\label{densitydec4}
\ee 
and thus
\be \label{transfer4}
{\bar\rho_{\chi_4} \over \rho_\phi} \sim 0.0166 \, g_Y^2\,\exp\!\left[{-\frac{\pi m_{\chi_4}^2}{2m_\phi^2\sqrt q_4}}\!\right]\,,
\ee
is significantly larger than that coming from each of the $\chi_{1,2,3}$ fields. The expressions in Eqs.~(\ref{dectime4},\ref{densitydec4},\ref{transfer4}) also hold for the $B$ gauge boson and $\Psi_4$ fermion, which have the same mass and decay width as $\chi_4$.



\subsection{Backreaction of $\chi$ particles on the inflaton}

As we will show here, the effective inflaton mass induced by the produced $\chi$ quanta,
\be\label{effem2}
m^2_{{\rm eff},\phi} = m_\phi^2 + \frac{1}{6}g^2_W\sum_{i=1}^3\langle\chi_i^2\rangle +
\frac{1}{2}g^2_Y\langle\chi_4^2\rangle\,,
\ee
does not modify significantly the inflaton oscillations. The $\chi$ variance induced at instant preheating can be computed exactly,
\begin{eqnarray} \nonumber
\langle\chi^2\rangle &\!=\!& \int_0^\infty \frac{k^2\,dk}{2\pi^2} \frac{n_k}{\omega_k(\tau)}\,e^{-\int \Gamma_\chi dt}  \\ \label{chi2}
&\!=\!&  \frac{m_\phi^2\sqrt q}{2\pi^3}\,Ae^A\Big(K_1(A)-K_0(A)\Big)\, \nonumber \\
&\!\times &\! \exp\!\left[-\frac{A}{2A_{\rm dec}}\!\right]\,\exp\!\left[-\frac{\pi m_\chi^2}{2m_\phi^2\sqrt q}\!\right] \, ,
\end{eqnarray}
which decays exponentially with time and never constitutes any danger for the rate of inflaton oscillations, e.g. at the time of
decay it is less than a few percent of $m^2_\phi$.


\subsection{Rate of Energy Transfer}

To calculate the rate of energy transfer via instant preheating, we must add the contributions from all 32 degress of freedom: 4 from scalars ($\chi_{1,2,3,4}$), 12 from gauge bosons ($W^{\pm},~W_3,~B$), and 16 from fermions ($\Psi_{1,2,3,4}$).
The fraction of transferred energy through $\chi_{1,2,3}$ scalars, $W^{\pm},~W_3$ gauge bosons, and $\Psi_{1,2,3}$ fermions (24 degrees of freedom in total) follows Eq.~(\ref{transfer}), while that through the $\chi_4$ scalar, $B$ gauge boson, and $\Psi_4$ fermion (8 degrees of freedom in total) is given by Eq.~(\ref{transfer4}).
Accordingly, the total energy transferred to (s)quarks via instant preheating will be $\rho_{\rm rel} = 24 {\bar \rho}_{\chi_1} + 8 {\bar \rho}_{\chi_4}$. For $g_Y \sim g_W \sim 0.6$, this results in
\be \label{ttransfer}
{\rho_{\rm rel} \over \rho_\phi} \sim 10.6 \%~~~({\rm per~zero-crossing}).
\ee
%
Note that this fraction is independent of the amplitude of oscillations. Therefore the energy density of the inflaton and quarks/squarks after $N$ oscillations of the inflaton field respectively are
\be \label{Nfraction}
\rho_\phi = 0.79^N \rho_0 \,, \hspace{1cm} \rho_{\rm rel} = (1-0.79^{N}) \rho_0 ,
\ee
where $\rho_0 \simeq \dot\phi_0^2/2$ is the inflaton energy density at the beginning of oscillations.

\begin{figure}[htb]
\begin{center}
\begin{tabular}{c}
\includegraphics[width=8.5cm]{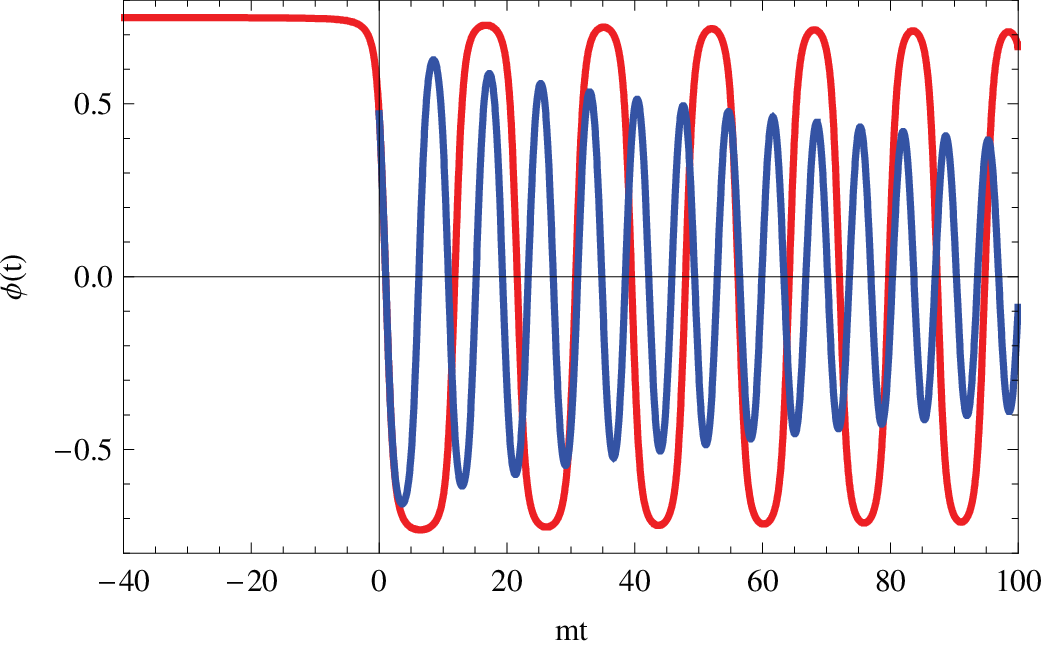}
\end{tabular}
\caption{The inflaton oscillations around the minimum of the potential after inflation. The red line corresponds to the inflaton evolution without including the particle production. The blue one takes into account the non-perturbative production of the $\chi$ quanta, and their perturbative decay to relativistic particles, which drain energy from the inflaton.}
\label{phitime}
\end{center}
\end{figure}


\section{Thermalization}

In this section we discuss thermalization of particles produced via instant preheating and establishment of thermal equilibrium among all of the MSSM degrees of freedom. This is when reheating completes and one can assign a reheating temperature to the universe.

\subsection{Thermalization of (s)quarks}

As we discussed, each time that the inflaton crosses the origin, about $10\%$ of its energy density is transferred into relativistic (s)quarks.
The produced (s)quarks do not have a thermal distribution, and the question is how long it takes for them to thermalize.

In thermal equilibrium, the number density and energy density of relativistic species are given by $n_{\rm thr} = (\zeta(3)/\pi^2) g_* T^3$ and $\rho_{\rm thr} = (\pi^2/30) g_* T^4$, respectively, where $g_*$ is the number of relativistic degrees of freedom and $T$ is the temperature. For a thermal distribution of (s)quarks, where $g_* = 135$, this results in $\rho_{\rm thr}^{1/4} \approx n_{\rm thr}^{1/3}$.

On the other hand, the number density and energy density of (s)quarks after the first inflaton zero-crossing are given by $n_{\rm rel} = 24 n_{\chi_1} + 8 n_{\chi_4}$ and $\rho_{\rm rel} = 24 {\bar \rho}_{\chi_1} + 8 {\bar \rho}_{\chi_4}$, where $n_{\chi_1},~\rho_{\chi_1},~n_{\chi_4},~\rho_{\chi_4}$ are given by Eqs.~(\ref{ndensity},\ref{transfer},\ref{ndensity4},\ref{transfer4}) respectively. We then find that $\rho^{1/4}_{\rm rel} \approx 1.6 n^{1/3}_{\rm rel}$. This implies that the number density of (s)quarks must increase by a mild factor in order for a thermal distribution to be established.

The main processes that increase the number density of particles are $2 \rightarrow 3$ scatterings of (s)quarks (with gluon exchange in the $t$-channel) that result in emission of gluons of energy $\sim \rho^{1/4}_{\rm rel}$~\cite{DS}. These processes are effective because the $SU(3)_C$ symmetry is not broken by the inflaton VEV, and hence the gluons remain massless at all times. Upon production, each gluon participates in the subsequent scatterings, which implies an increasingly faster rate for the process. It is therefore suggested that the rate for the $2 \rightarrow 3$ scatterings can be considered as the rate for thermalization of colored fields. This is estimated to be~\cite{DS}
\be \label{thermalrate}
\Gamma^C_{\rm thr} \sim \alpha_C {n_{\rm rel} \over \rho^{1/2}_{\rm rel}} ,
\ee
where $\alpha_C$ is the $SU(3)_C$ fine structure constant.

After using the relations $\rho_{\rm rel} \approx (1.6)^{4} n^{4/3}_{\rm rel}$ and $\rho_{\rm rel} \sim 0.1 \rho_\phi$, see Eq.~(\ref{ttransfer}), we find that $\Gamma^C_{\rm thr} \gg m_\phi$.\footnote{A more careful treatment that takes various effects (such as multiple scatterings) into account results in a rate that is somewhat different from that in Eq.~(\ref{thermalrate}) (we would like to thank Guy Moore for clarifying this issue). This however does not affect our results, since the more precise value for $\Gamma_{\rm thr}$ will still be much larger.} This implies that a thermal bath consisting of the colored particles indeed forms much earlier than the next zero-crossing of the inflaton. It initially carries $\sim 10\%$ of the inflaton energy density, which results in a temperature $T \sim 0.2 (m_\phi \phi_0)^{1/2}$. The energy density, hence temperature, of the thermal bath grows after each inflaton zero-crossing as more (s)quarks are produced via instant preheating.

Because of the large Yukawa coupling of the top quark, $H_u$ and its Higgsino partner ${\tilde H}_u$ are also brought into thermal equilibrium with the colored particles. However, the thermal bath does not contain (s)leptons because they interact with (s)quarks only via $SU(2)_W \times U(1)_Y$ gauge interactions, which are suppressed due to the breaking of this symmetry by the large inflaton VEV.\footnote{The electroweak symmetry is restored over a short period each time that the inflaton crosses the origin. However, this time is too short for the gauge interactions to bring the (s)leptons into equilibrium with the thermal bath.}

Therefore combination of instant preheating and thermalization via $SU(3)_C$ gauge interactions provides a very efficient mechanism for transferring the energy from the inflaton to a thermal bath of colored particles. The transfer of energy results in a continuous decrease in the amplitude of oscillations ${\hat \phi}$, while temperature of the thermal bath $T$ keeps increasing. As a result, this thermal bath will dominate the energy density after several oscillations.


\subsection{Thermal masses}

Production of $\chi$ particles is affected by their mass $m_\chi$ according to Eqs.~(\ref{ndensity},\ref{ndensity4}). For $m^2_\chi > 0$ the number density of $\chi$ is exponentially suppressed, while $m^2_\chi < 0$ results in an enhancement of $n_\chi$. Particle production occurs due to violation of adiabaticity in $\omega_k$, see Eq.~(\ref{eigenmodes}), each time that the inflaton crosses the origin. A positive $m^2_\chi > 0$ increases $\omega_k$ and decreases ${\dot \omega}_k$. As a result, the adiabaticity condition is violated for a shorter period of time, as compared with the $m_\chi = 0$ case, which suppresses particle production. A negative $m^2_\chi$ has the opposite effect and enhances particle production.

An important contribution to $m_\chi$ is the effective mass induced by $\chi$ coupling to fields that are in thermal equilibrium.
This results in thermal masses, which can be derived from Eqs.~(\ref{scalint1},\ref{scalint2}):
$$V \supset {g^2_Y \over 4} (\chi^2_4 - \sum_{j=1}^{3}{\chi^2_j}) \left[{1 \over 6} (\vert {\tilde Q}_i\vert^2 - 4 \vert{\tilde u}_i\vert^2 + 2 \vert{\tilde d}_i\vert^2) + {1 \over 2} \vert H_{u} \vert^2 \right] \,. $$
Here we have only kept those fields that are in thermal equilibrium, namely the squarks and $H_u$.

In thermal equilibrium, for any complex scalar $\Phi$ we have $\langle \Phi^2 \rangle = T^2/6$ (see, for example~\cite{ce}). This results in
\begin{eqnarray} \label{meff1}
m^2_{\chi_1,{\rm eff}} = m^2_{\chi_2,{\rm eff}} = m^2_{\chi_3,{\rm eff}} = -\frac{g_Y^2 T^2}{12} \,,~
m^2_{\chi_4,{\rm eff}} = {g^2_Y T^2 \over 12}  \, . \nonumber \\
\,
\end{eqnarray}
At first, the tachyonic thermal masses for $\chi_{1,2,3}$ may look confusing. As discussed in Ref.~\cite{ce}, the individual contributions of scalar fields to the thermal mass of another scalar (induced via $D$-term interactions) can be positive or negative. If all scalars are in thermal equilibrium, we then have $m^2_{\rm eff} > 0$. However, in our case only the squarks and $H_u$ are in equilibrium. The other scalars have electroweak interactions with the thermal bath, which are suppressed by the large inflaton VEV. The electroweak symmetry is restored for a very short period of time when the inflaton crosses the origin. However, the time is too short to bring the inflaton and $\chi$'s into equilibrium.\footnote{Continuous transfer of the inflaton energy to the thermal bath via instant preheating lowers the amplitude of its oscillations. Eventually, when the amplitude becomes comparable with the temperature, electroweak interactions become
efficient and all degrees of freedom will come to equilibrium. At this point we will have $m^2_{\chi,{\rm eff}} > 0$, as expected.}

We note that the tachyonic thermal masses of $\chi_{1,2,3}$ are important only at the inflaton zero crossing. As the inflaton rolls back to large VEV's, its contribution to $\omega_k$ dominates that from the thermal bath, which prevents the latter from inducing instabilities.

We also note that $m_{\chi,\rm eff}$ is zero initially (i.e. before the first burst of particle production) but increases each time that the inflaton crosses the origin and produces squarks via instant preheating. We can equate the energy density in these relativistic particles with that of the produced $\chi$ particles,
\be
\rho_{\rm rel} = 0.1055\,\rho_\phi = \frac{\pi^2}{30}g_*\,T^4\,,
\ee
where $g_* = 172.5$ is the number of thermal degrees of freedom (i.e., the colored particles plus $H_u$ and ${\tilde H}_u$). Thus
\be
|m_\chi^2| \simeq \frac{g^2_Y}{12}\left(\frac{3.2}{\pi^2g_*}\right)^{1/2} \rho_\phi^{1/2}\,,
\ee
or
\be
\frac{\pi |m_\chi^2|}{2m_\phi^2\sqrt q_4} \simeq \frac{g_Y}{12}\left(\frac{3.2}{g_*}\right)^{1/2} \sqrt{\rho_\phi \over \rho_0}\,.
\ee
Writing these masses in terms of the initial inflaton energy we arrive at the expression
\be\label{decrel}
{\rho_{\rm rel} \over \rho_\phi} \sim 0.058\,\exp\!\left[\frac{\sqrt{\rho_\phi/\rho_0}}{158}\!\right]+
0.048\,\exp\!\left[-\frac{\sqrt{\rho_\phi/\rho_0}}{158}\!\right]\,,
\ee
for the inflaton energy converted into relativistic species every half oscillation, $m_\phi \Delta t \sim 5$.


\subsection{Coupled evolution of the relative energy densities}

We can thus describe the evolution after inflation as a coupled system in which the inflaton energy density is transferred to the $\chi$ fields, whose particles promptly decay into relativistic particles that immediately thermalize due to the rapid rate of strong interactions. This leads to a set of coupled differential equations,
\bea \label{coupled}
\dot\rho_\phi &\!=\!& - \left(\Gamma_\phi^{(1)} + \Gamma_\phi^{(4)}\right) \,\rho_\phi\,\exp\!\left[\frac{\sqrt{\rho_\phi/\rho_0}}{158}\!\right]
\, \\
\dot\rho_{\chi_1} &\!=\!& \Gamma_\phi^{(1)}\,\rho_\phi\,\exp\!\left[\frac{\sqrt{\rho_\phi/\rho_0}}{158}\!\right] -
\Gamma_{\chi_1}\,\sqrt{\rho_\phi}\,\rho_{\chi_1}\\
\dot\rho_{\chi_4} &\!=\!& \Gamma_\phi^{(4)}\,\rho_\phi\,\exp\!\left[-\frac{\sqrt{\rho_\phi/\rho_0}}{158}\!\right] -
\Gamma_{\chi_4}\,\sqrt{\rho_\phi}\,\rho_{\chi_4}\\
\dot\rho_{\rm rel} &\!=\!& \Gamma_{\chi_1}\,\sqrt{\rho_\phi}\,\rho_{\chi_1} + \Gamma_{\chi_4}\,\sqrt{\rho_\phi}\,\rho_{\chi_4}
\eea
where $\Gamma_\phi^{(1)} \sim 0.0116 \, m_\phi$ denotes the ``effective" decay rate due to the loss of energy into $\chi_{1,2,3}$ (as well as $W^{\pm},~W_3$ and $\Psi_{1,2,3}$) particles, $\Gamma_\phi^{(4)}\sim 0.0096 \, m_\phi$ denotes the ``effective'' decay rate due to the loss of energy into into $\chi_4$ (as well as $B$ and $\Psi_4$) particles, see Eq.~(\ref{decrel}). Also, $\Gamma_{\chi_{1(4)}}\sim 1.2(1.7)\times10^{10} \, m_\phi$ denote the ``effective" decay rate of $\chi_{1,2,3(4)}$ particles into (s)quarks, in units of $\sqrt{\rho_\phi} \propto \phi$. The latter decay rate is so large that there is simply no time for building up particle occupation numbers (and thus resonant production like in parametric resonance) of $\chi$ fields.

\begin{figure}[htb]
\begin{center}
\begin{tabular}{c}
\includegraphics[width=8.5cm]{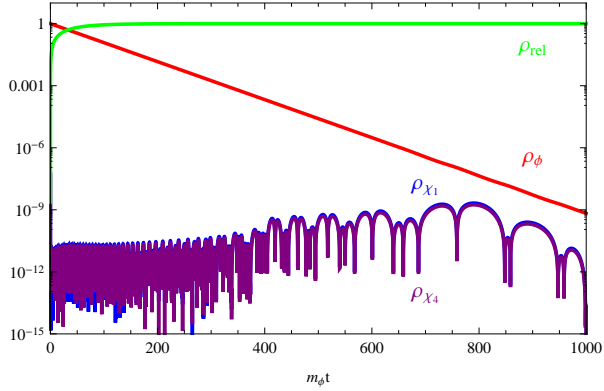}
\end{tabular}
\caption{The evolution of the different energy densities, normalized to the initial energy density in the inflaton $\rho_0$, as a function of time (in units of $m_\phi$) according to Eqs.~(42-45). Note that most of the energy of the inflaton, $\rho_\phi$, is transferred to relativistic particles after a few inflaton oscillations.}
\label{coupledevo}
\end{center}
\end{figure}

We have solved the set of differential equations in Eqs.~(\ref{coupled},43,44,45) for initial conditions $\rho_\phi(0)/\rho_0 = 1$, $\rho_{\chi_1}(0)/\rho_0 = 0$, $\rho_{\chi_4}(0)/\rho_0 = 0$ and $\rho_{\rm rel}(0)/\rho_0 = 0$, and plotted the results in Fig.~\ref{coupledevo}.

We have also performed the integration of the above coupled equations assuming that the thermal masses are absent (ignoring the exponential terms in Eqs.(42-45)), and we see no appreciable difference.

It is seen from Fig.~3 that $99\%$ of the inflaton energy density is transferred within 20 or so oscillations, which is in agreement with the expression in Eq.~(\ref{Nfraction}). Therefore, we conclude that the inflaton energy decays into a thermal bath (consisting of colored particles, $H_u$ and ${\tilde H}_u$) within ${\cal O}(20)$ oscillations, long before the expansion of the universe has diluted its energy. As a consequence, we can assume that reheating is very efficient in this type of theories.


\subsection{Reheating temperature of the universe}

Even though the colored particles thermalize at a rate $\Gamma^C_{\rm thr}$, see Eq.~(\ref{thermalrate}), which is much faster than $m_\phi$, all of the MSSM degrees of freedom do not reach equilibrium as quickly. Full thermal equilibrium is achieved when all of the gauge interactions become efficient~\cite{AM-reh}. As mentioned earlier, the electroweak interactions are initially suppressed, because the large VEV of the $LLe$ inflaton breaks the $SU(2)_W \times U(1)_Y$ symmetry completely.

The amplitude of inflaton oscillations ${\hat \phi}$ decreases as the energy is continuously transferred via instant preheating. Eventually, we reach a point when the VEV-induced masses of the $\chi$ scalars, electroweak gauge bosons and $\Psi$ fermions, see Eqs.~(\ref{scalcoupl},\ref{gaugcoupl},\ref{fermcoupl}), is smaller than the average kinetic energy of particles in thermal equilibrium at all times, i.e.
\be \label{thrm}
{g_W {\hat \phi} \over \sqrt{6}} , ~ {g_Y {\hat \phi} \over \sqrt{2}} \lsim 3 T.
\ee
After using Eq.~(\ref{Nfraction}), relation $\rho_{\rm rel} \approx \rho_0 = (\pi^2/30) g_* T^4$ (where we take $g_* = 172.5$), and the fact that ${\hat \phi} \propto \rho^{1/2}_\phi$, we find that it takes $N \sim {\cal O}(100)$ oscillations for this to occur.

At this point all of the MSSM particles are kinematically available to the thermal bath and all of the gauge interactions act efficiently. In particular, the $2 \rightarrow 2$ scatterings via electroweak gauge interactions happen at a rate $\sim \alpha^2_{EW} T$, with $\alpha_{EW}$ denoting the electroweak gauge fine structure constant, which is much larger than $m_\phi$. As a result, scatterings of (s)quarks will bring (s)leptons and electroweak gauge/gaugino particles into thermal equilibrium, and destroys the residual inflaton condensate, very rapidly.

Therefore, we conclude that all of the MSSM degrees of freedom thermalize and reheating completes within ${\cal O}(100)$ oscillations. Note that due to the hierarchy between $H_{\rm inf} \sim 10^{-3} m_\phi$, this happens within a single Hubble time after the end of inflation.
One can then assign a reheat temperature $T_{\rm rh}$ to the universe at this moment, which is given by
%
%
%
\be \label{Trh}
T_{\rm rh} = \left(\frac{30}{\pi^2 g_*}\right)^{1/4} \rho_0^{1/4} \simeq 2 \times 10^8 ~ {\rm GeV} \, ,
\ee
where we have used $g_* = 228.75$ (all degrees of freedom in MSSM) and $\rho_{0} = (4/15)m_{\phi}^{2}\phi_{0}^{2}$, see Eq.~(\ref{2ndinf}). We note that this reheating temperature is compatible with the Big Bang Nucleosynthesis bounds on gravitino production for a gravitino mass $m_{3/2} \gsim {\cal O}({\rm TeV})$. It is also sufficiently high that various mechanisms of baryogenesis may be invoked to generate the observed baryon asymmetry of the universe.

We can summarize the relevant time scales in reheating after MSSM inflation as follows:
\begin{itemize}
\item{$\tau_{\rm osc} = 2 \pi m^{-1}_\phi$: the period of inflaton oscillations, which separates successive bursts of non-perturbative particle
production.}
\item{$\tau_\phi \sim {\cal O}(20) \tau_{\rm osc}$: the time scale for efficient transfer of the inflaton energy to relativistic particles via
instant preheating, which can be considered as the effective inflaton decay lifetime,}
\item{$\tau_{\rm thr} \sim {\cal O}(100) \tau_{\rm osc}$: the time scale for all observable degrees of freedom to come into equilibrium, which
represents the end of reheating,}
\item{$\tau_{\rm exp} \sim 1000 m^{-1}_\phi$: the Hubble time right after the end of inflation, which sets the expansion rate of the
universe.}
\end{itemize}
It is important to note that the time scales involved in microphysical processes $\tau_{\rm osc},~\tau_\phi,~\tau_{\rm thr}$ are nicely separated and they are all smaller than one Hubble time $\tau_{\rm exp}$. This has simplified our analysis of reheating after MSSM inflation.

\section{Conclusion}

In this paper we have studied an inflationary model that naturally gives rise to a successful and efficient reheating of the observable sector and excites the MSSM degrees of freedom.
As a result, it naturally explains the creation of matter that makes the structures we see in the universe today.\footnote{An alternative scenario is Higgs inflation~\cite{BS2008}, although reheating seems to be less efficient in this case~\cite{GBFR2009}, due to the large VEV of the Higgs-inflaton after inflation.} We focused on the particular case where the $LLe$ flat direction is the inflaton. We identified the inflaton couplings to the scalars, gauge bosons and fermions and their couplings to other fields that are not coupled to the inflaton. We showed that the the former are produced non-perturbatively every time that the inflaton passes through the minimum of its potential. As the inflaton rolls back the potential, they become heavy, which results in their very fast perturbative decay to the latter.

The combination of these effects results in a very efficient transfer of the inflaton energy to relativistic particles via instant preheating, which drains about $20\%$ of the energy density in a single oscillation. After ${\cal O}(20)$ oscillations virtually all of the inflaton energy density is transferred into relativistic particles, and full thermal equilibrium in the observable sector will be achieved after ${\cal O}(100)$ oscillations. This takes less than one Hubble time, which provides by far the most efficient reheating of the universe with the observed degrees of freedom.

This study also lays down the foundations for reheating in inflationary models where the inflaton couples to all the degrees of freedom with the Standard Model gauge couplings. Our analysis holds in general at a qualitative level, and the quantitative results can be applied with proper modifications.

\section{Acknowledgement}

We would like to thank Kari Enqvist, Jose Ram\'on Espinosa, Daniel G. Figueroa, Andrei Linde and Guy Moore for helpful discussions.
JGB thanks the Institute de Physique Th\'eorique de l'Universit\'e de Gen\`eve for their generous hospitality during his sabbatical in Geneva. This work is supported by the University of New Mexico Office of Research, the Spanish MICINN under project AYA2009-13936-C06-06, the C.A.M. project HEPHACOS (Ref. S2009/ESP-1473), the EU FP6 Marie Curie Research and Training Network ``UniverseNet" (MRTN-CT-2006-035863), and the Academy of Finland through grant 114419.


\section{Appendix}

Here we present the essential expressions that are needed to derive the inflaton couplings to scalars, gauge bosons, and fermions as well as the decay widths of the latter. The starting point is the MSSM Lagrangian, which can be found, e.g., in~\cite{HK}.


\subsection{Scalar interactions}
\label{DWOS}

Couplings of the scalar fields to the inflaton are governed by the electroweak $D$-terms:
\begin{eqnarray} \label{scalint1}
& V_D & \supset {1 \over 2} g^2_W (D^2_1 + D^2_2 + D^2_3) + {1 \over 2} g^2_Y D^2_Y \, ,
\end{eqnarray}
where
\begin{equation} \label{D}
D_a = \sum_{i} {\Phi^{\dagger}_i T^a \Phi_i} ~ ~ ~ , ~ ~ ~ D_Y = \sum_{i} {{1 \over 2} y_i \Phi^{\dagger}_i \Phi_i} .
\end{equation}
Here $\Phi_i$ denotes a scalar field with hypercharge $y_i$, and $T^a$ are the $SU(2)_W$ generators.

Writing ${\tilde L}_1,~{\tilde L}_2,~{\tilde e}_3$ in terms of the the inflaton $\phi$ and $\chi$'s, see Eq.~(\ref{scaldef}), the electroweak $D$-terms read
\begin{eqnarray} \label{scalint2}
D_1 & = & {\phi \chi_1 \over \sqrt{6}} - {\chi_1 \chi_4  \over 2 \sqrt{3}} + {{\tilde Q}^{\dagger}_i T^1 {\tilde Q}_i} + H^{\dagger}_{u,d} T^1 H_{u,d} + {\tilde L}^{\dagger}_3 T^1 {\tilde L}_3 \,  \nonumber \\
D_2 & = & {\phi \chi_2 \over \sqrt{6}} - {\chi_2 \chi_4 \over 2 \sqrt{3}} + {{\tilde Q}^{\dagger}_i T^2 {\tilde Q}_i} + H^{\dagger}_{u,d} T^2 H_{u,d} + {\tilde L}^{\dagger}_3 T^2 {\tilde L}_3 \,  \nonumber \\
D_3 & = & {\phi \chi_3 \over \sqrt{6}} - {\chi_3 \chi_4 \over 2 \sqrt{3}} + {{\tilde Q}^{\dagger}_i T^3 {\tilde Q}_i} + H^{\dagger}_{u,d} T^3 H_{u,d} + {\tilde L}^{\dagger}_3 T^3 {\tilde L}_3 \,  \nonumber \\
D_Y & = & {\phi \chi_4 \over \sqrt{2}} + {\chi^2_4 - \chi^2_1 - \chi^2_2 - \chi^2_3 \over 4} \, \nonumber \\
& + & {{1 \over 6} (\vert {\tilde Q}_i\vert^2 - 4 \vert{\tilde u}_i\vert^2 + 2 \vert{\tilde d}_i\vert^2}) + (\vert {\tilde e}_1 \vert^2 + \vert {\tilde e}_2 \vert)^2) \, \nonumber \\
& + & {1 \over 2} (\vert H_{u} \vert^2 - \vert H_d \vert^2 - \vert {\tilde L}_3 \vert^2) \, , \nonumber \\
& & \,
\end{eqnarray}
where $i$ denotes the family index of squarks (we have omitted color indices for simplicity).

The terms that are relevant for the decay of $\chi$'s to squarks, Higgs particles, and ${\tilde L}_3,~{\tilde e}_1,~{\tilde e}_2$ sleptons are
%
\begin{eqnarray} \label{chidec}
V & \supset & {g_W^2 \phi \over \sqrt{6}} ~ \chi_1 \Big({\tilde Q}^{\dagger}_{i} T^1 {\tilde Q}_{i} + H^{\dagger}_{u} T^1 H_{u} + H^{\dagger}_d T^1 H_d + {\tilde L}^{\dagger}_3 T^1 {\tilde L}_3 \Big) \,  \nonumber \\
& + & {g_W^2 \phi \over \sqrt{6}} ~ \chi_2 \Big({\tilde Q}^{\dagger}_{i} T^2 {\tilde Q}_{i} + H^{\dagger}_{u} T^2 H_{u} + H^{\dagger}_d T^2 H_d + {\tilde L}^{\dagger}_3 T^2 {\tilde L}_3 \Big) \,  \nonumber \\
& + & {g_W^2 \phi \over \sqrt{6}} ~ \chi_3 \Big({\tilde Q}^{\dagger}_{i} T^3 {\tilde Q}_i  +  H^{\dagger}_{u} T^3 H_u + H^{\dagger}_d T^3 H_d + {\tilde L}^{\dagger}_3 T^3 {\tilde L}_3 \Big) \,  \nonumber \\
& + & {g_Y^{ 2} \phi \over \sqrt{2}} ~ \chi_4 \Big[{1 \over 6} (\vert {\tilde Q}_{i} \vert^2 - 4 \vert{\tilde u}_i \vert^2 + 2 \vert {\tilde d}_i \vert^2) + (\vert {\tilde e}_1 \vert^2 + \vert {\tilde e}_2 \vert)^2) \, \nonumber \\
& + & {1 \over 2} (\vert H_{u}\vert^2 - \vert H_{d}\vert^2 - \vert {\tilde L}_3 \vert^2) \Big] \, .
\end{eqnarray}
An interaction term $\sigma \chi \phi^* \phi$ between $\chi$ and a massless scalar $\phi$ results in a decay rate $\sigma^2/16 \pi M$, where $\sigma$ is a coupling of dimension mass and $M$ is the mass of $\chi$.

\subsection{Gauge field interactions}
\label{DWOGF}

Couplings of the inflaton to the gauge fields are obtained from the flat direction kinetic terms
\begin{eqnarray} \label{kin}
{\cal L} \supset (D^\mu {\tilde L}_1)^{\dagger} (D_\mu {\tilde L}_1) +  (D^\mu {\tilde L}_2)^{\dagger} (D_\mu {\tilde L}_2) + (D^\mu {\tilde e}_3)^{\dagger} (D_\mu {\tilde e}_3) \, , \nonumber \\
\,
\end{eqnarray}
where
\begin{eqnarray} \label{der}
D_\mu {\tilde L}_1 & = & (\partial_\mu + {i \over 2} g_Y B_\mu - i g_W \sum_{a=1}^{3}{W_{a,\mu} T^a})
{\tilde L}_1 \, \nonumber \\
D_\mu {\tilde L}_2 & = & (\partial_\mu + {i \over 2} g_Y B_\mu - i g_W \sum_{a=1}^{3}{W_{a,\mu} T^a})
{\tilde L}_2 \, \nonumber \\
D_\mu {\tilde e}_3 & = & (\partial_\mu - i g_Y B_\mu ) {\tilde e}_3 \, .
\end{eqnarray}
Here $W_{1,\mu},~W_{2,\mu},~W_{3,\mu}$ and $B_\mu$ are the gauge fields of $SU(2)_W$ and $U(1)_Y$ respectively.

The gauge fields decay to (s)quarks, the Higgses and Higgsinos, and the $L_3,~e_1,e_2$ (s)leptons. The interaction terms that are relevant for the decay to the scalars are given by
\begin{eqnarray} \label{gaugscint}
& {\cal L} & \supset {i g_W \over \sqrt{2}} ({\tilde Q}^*_{u,i} W^{+,\mu} {\partial}_\mu {\tilde Q}_{d,i} - {\partial}_\mu {\tilde Q}^{*}_{u,i} W^{+,\mu} {\tilde Q}_{d,i}) + {\rm h.c.} \, \nonumber \\
& + & {i g_W \over \sqrt{2}} ({\tilde L}^{1 *}_{3} W^{+,\mu} {\partial}_\mu {\tilde L}^2_{3} - {\partial}_\mu{\tilde L}^{1 *}_{3} W^{+,\mu} {\tilde L}^2_{3}) + {\rm h.c.} \, \nonumber \\
& + & {i g_W \over \sqrt{2}} ({H}^{1 *}_{u} W^{+,\mu} {\partial}_\mu {H}^2_{u} - {\partial}_\mu{H}^{1 *}_{u} W^{+,\mu} {H}^2_{u}) + {\rm h.c.} \, \nonumber \\
& + & {i g_W \over \sqrt{2}} ({H}^{1 *}_{d} W^{+,\mu} {\partial}_\mu {H}^2_{d} - {\partial}_\mu{H}^{1 *}_{d} W^{+,\mu} {H}^2_{d}) + {\rm h.c.} \, \nonumber \\
& + & {i g_W \over 2} ({\tilde Q}^{*}_{u,i} W^{\mu}_3 {\partial}_\mu {\tilde Q}_{u,i} - {\tilde Q}^*_{d,i} W^{\mu}_3 {\partial}_\mu {\tilde Q}_{d,i}) + {\rm h.c.} \, \nonumber \\
& + & {i g_W \over 2} ({\tilde L}^{1 *}_{3} W^{\mu}_3 {\partial}_\mu {\tilde L}^1_{3} - {\tilde L}^{2 *}_{2} W^{\mu}_3 {\partial}_\mu {\tilde L}^2_{3}) + {\rm h.c.} \, \nonumber \\
& + & {i g_W \over 2} ({H}^{1 *}_{u} W^{\mu}_3 {\partial}_\mu {H}^1_{u} - {H}^{2 *}_{u} W^{\mu}_3 {\partial}_\mu {H}^2_{u}) + {\rm h.c.} \, \nonumber \\
& + & {i g_W \over 2} ({H}^{1 *}_{d} W^{\mu}_3 {\partial}_\mu {H}^1_{d} - {H}^{2 *}_{d} W^{\mu}_3 {\partial}_\mu {H}^2_{d}) + {\rm h.c.} \, \nonumber \\
& + & {i g_Y \over 6} ({\tilde Q}^*_{i} B^{\mu} {\partial}_\mu {\tilde Q}_{i} - 4 {\tilde u}^*_{i} B^{\mu} {\partial}_\mu {\tilde u}_{i} + 2 {\tilde d}^*_{i} B^{\mu} {\partial}_\mu {\tilde d}_{i}) + {\rm h.c.} \, \nonumber \\
& + & {i g_Y \over 2} ({H}^{*}_{u} B^{\mu} {\partial}_\mu {H}_{u} - {H}^{*}_{d} B^{\mu} {\partial}_\mu {H}_{d} - {\tilde L}^*_{3} B^{\mu} {\partial}_\mu {\tilde L}_{3}) + {\rm h.c.} \, \nonumber \\
& + & {i g_Y \over 2} ( 2 {\tilde e}^*_{1} B^{\mu} {\partial}_\mu {\tilde e}_{1} + 2 {\tilde e}^*_{2} B^{\mu} {\partial}_\mu {\tilde e}_{2} + {\rm h.c.}) \, .
\end{eqnarray}
For a massless scalar $\phi$, the rates for the decay of $SU(2)_W$ and $U(1)_Y$ gauge fields to $\phi^* \phi$ pair are $g^2_W M/96 \pi$ and $g^2_Y y^2 M/96 \pi$ respectively. Here $M$ is the gauge fields mass and $y$ is the hypercharge of $\phi$.

The relevant terms for the decay of gauge fields to fermions are given by
\begin{eqnarray} \label{gaugfeint}
& {\cal L} & \supset - {i g_W \over \sqrt{2}} ({\bar {\Psi}}_{u,i} W^{+,\mu} \gamma_\mu P_L {\Psi}_{d,i} + {\bar {\Psi}}^1_{l,3} W^{+,\mu} \gamma_\mu P_L {\Psi}^2_{l,3}) + {\rm h.c.} \, \nonumber \\
& - & {i g_W \over \sqrt{2}} ({\bar {\Psi}}^1_{H_u} W^{+,\mu} \gamma_\mu P_L {\Psi}^2_{H_u} + {\bar {\Psi}}^1_{H_d} W^{+,\mu} \gamma_\mu P_L {\Psi}^2_{H_d}) + {\rm h.c.} \, \nonumber \\
& - & {i g_W \over 2} ({\bar {\Psi}}_{u,i} W^{\mu}_3 \gamma_\mu P_L {\Psi}_{u,i} - {\bar {\Psi}}_{d,i} W^{\mu}_3 \gamma_\mu P_L {\Psi}_{d,i}) \, \nonumber \\
& - & {i g_W \over 2} ({\bar {\Psi}}^1_{l,3} W^{\mu}_3 \gamma_\mu P_L {\Psi}^1_{l,3} - {\bar {\Psi}}^2_{l,3} W^{\mu}_3 \gamma_\mu P_L {\Psi}^2_{l,3}) \, \nonumber \\
& - & {i g_W \over 2} ({\bar {\Psi}}^1_{H_u} W^{\mu}_3 \gamma_\mu P_L {\Psi}^1_{H_u} - {\bar {\Psi}}^2_{H_u} W^{\mu}_3 \gamma_\mu P_L {\Psi}^2_{H_u}) \, \nonumber \\
& - & {i g_W \over 2} ({\bar {\Psi}}^1_{H_d} W^{\mu}_3 \gamma_\mu P_L {\Psi}^1_{H_d} - {\bar {\Psi}}^2_{H_d} W^{\mu}_3 \gamma_\mu P_L {\Psi}^2_{H_d}) \, \nonumber \\
& - & {i g_Y \over 6} ({\bar {\Psi}}_{u,i} B^{\mu} \gamma_\mu P_L {\Psi}_{u,i} + {\bar {\Psi}}_{d,i} B^{\mu} \gamma_\mu P_L {\Psi}_{d,i}) \, \nonumber \\
& - & {i g_Y \over 2} ({\bar {\Psi}}^1_{H_u} B^{\mu} \gamma_\mu P_L {\Psi}^1_{H_u} + {\bar {\Psi}}^2_{H_u} B^{\mu} \gamma_\mu P_L {\Psi}^2_{H_u}) \, \nonumber \\
& - & {i g_Y \over 2} (- {\bar {\Psi}}^1_{H_d} B^{\mu} \gamma_\mu P_L {\Psi}^1_{H_d} - {\bar {\Psi}}^2_{H_d} B^{\mu} \gamma_\mu P_L {\Psi}^2_{H_d}) \, \nonumber \\
& - & {i g_Y \over 2} (- {\bar {\Psi}}^1_{l,3} B^{\mu} \gamma_\mu P_L {\Psi}^1_{l,3} - {\bar {\Psi}}^2_{l,3} B^{\mu} \gamma_\mu P_L {\Psi}^2_{l,3}) \, \nonumber \\
& - & {i g_Y \over 6} (- 4 {\bar {\Psi}}_{u,i} B^{\mu} \gamma_\mu P_R {\Psi}_{u,i} + 2 {\bar {\Psi}}_{d,i} B^{\mu} \gamma_\mu P_R {\Psi}_{d,i}) \,  \nonumber\\
& - & {i g_Y \over 2} (2 {\bar {\Psi}}^2_{l,1} B^{\mu} \gamma_\mu P_R {\Psi}^2_{l,1} + 2 {\bar {\Psi}}^2_{l,2} B^{\mu} \gamma_\mu P_R {\Psi}^2_{l,2}) \, .
\end{eqnarray}
Here $P_L \equiv (1 + \gamma_5)/2$ and $P_R \equiv (1 - \gamma_5)/2$ are the left- and right-chiral projection operators respectively. $\Psi_u,~\Psi_d$ are Dirac spinors representing the up- and down-type quarks respectively
\begin{equation} \label{uddirac}
{\Psi}_{u,i} = \left(\begin{array}{cc} Q_{u,i} \\ - i \sigma_2 u^*_i \end{array}\right)~~ , ~~ {\Psi}_{d,i} = \left(\begin{array}{cc} Q_{d,i} \\ - i \sigma_2 d^*_i \end{array}\right).
\end{equation}
${\Psi}^1_{l}$ and $\Psi^2_{l}$ are Dirac spinors representing the neutrinos and charged leptons respectively (superscripts on $L_i$ denote the weak isospin component)
\begin{equation} \label{leptondirac}
{\Psi}^1_{l,i} = \left(\begin{array}{cc} L^1_i \\ 0 \end{array}\right)~~ , ~~ {\Psi}^2_{l,i} = \left(\begin{array}{cc} L^2_i \\ - i \sigma_2 e^*_i \end{array}\right),
\end{equation}
and
\begin{equation} \label{Higgdirac}
{\Psi}^1_{H_u} = \left(\begin{array}{cc} {\tilde H}^1_u \\ 0 \end{array}\right)~~ , ~~ {\Psi}^2_{H_u} =\left(\begin{array}{cc} {\tilde H}^2_u \\ 0 \end{array}\right).
\end{equation}
For a massless fermion $\psi$, the rates for the decay of $SU(2)_W$ and $U(1)_Y$ gauge fields to ${\bar \psi} \psi$ pair are $g^2_W M/48 \pi$ and $g^2_Y y^2 M/48 \pi$ respectively. Here $M$ is the gauge field mass and $y$ is the hypercharge of $\psi$.

\subsection{Fermion interactions}
\label{DWOF}

The inflaton couplings to the fermions are found from the following part of the Lagrangian
\begin{eqnarray} \label{fermlag}
&&{\cal L} \supset \sqrt{2} g_W \sum_{i=1}^{3}{\Big[ {\tilde L}^{\dagger}_1 {\tilde W}_i^t T^i (i \sigma_2 L_1) + {\tilde L}^{\dagger}_2 {\tilde W}_i^t T^i (i \sigma_2 L_2) \Big] }  \, \nonumber \\
&&+ \sqrt{2} g_Y \Big[ {\tilde e}^{\dagger}_3 {\tilde B}^t (i \sigma_2 e_3) - {1 \over 2} {\tilde L}^{\dagger}_1 {\tilde B}^t  (i \sigma_2 L_1) - {1 \over 2} {\tilde L}^{\dagger}_2 {\tilde B}^t (i \sigma_2 L_2) \Big] \,  \nonumber \\
&&+  {\rm h.c.} \,
\end{eqnarray}
Here ${\tilde W}_1,~{\tilde W}_2,~{\tilde W}_3$ and ${\tilde B}$ are the gauginos of $SU(2)_W$ and $U(1)_Y$ respectively. Superscript $t$ denotes transposition, and $\sigma_2$ is the second Pauli matrix.

The field content of $L_1,~L_2,e_3$ is
\begin{eqnarray} \label{ferm}
{L}_1 = \left(\begin{array}{ll}
\psi_1 \\ \psi_2 \end{array}\right), ~~
{L}_2 = \left(\begin{array}{ll} \psi_3 \\ \psi_4 \end{array}\right),
~~ {e}_3 = \psi_5 ,
\end{eqnarray}
We find the following Dirac spinors from pairing $\psi$'s and gauginos (which are left-handed Weyl spinors)
\begin{eqnarray} \label{fermdef}
{\Psi}_1 = & \left(\begin{array}{cc}
{\psi}_2 \\  \\ -i \sigma_2 {\tilde W}^{+*} \end{array}\right) & \, \nonumber \\
\, \nonumber \\
{\Psi}_2 = & \left(\begin{array}{cc} {\psi}_3 \\  \\ -i \sigma_2 {\tilde W}^{-*} \end{array}\right) & \, \nonumber \\
\, \nonumber \\
{\Psi}_3 = & \left(\begin{array}{cc} \Big({{\psi}_1 - {\psi}_4 \over \sqrt{2}}\Big) \\  \\ -i \sigma_2 {\tilde W}^*_3  \end{array}\right) & \, \nonumber \\
\, \nonumber \\
{\Psi}_4 = & \left(\begin{array}{cc} {\tilde B} \\ \\ \sqrt{2 \over 3} (-i \sigma_2) \Big({{\psi}_5 - {1 \over 2} {\psi}_1 - {1 \over 2} {\psi}_4}\Big)^*  \end{array} \right) & \, ,
\end{eqnarray}
where
\begin{eqnarray} \label{wino}
{\tilde W}^+ = {{\tilde W}_1 - i {\tilde W}_2 \over \sqrt{2}} ~ ~ ~ , ~ ~ ~ {\tilde W}^{-} = {{\tilde W}_1 + i {\tilde W}_2 \over \sqrt{2}} \, , \nonumber \\
\,
\end{eqnarray}
are the supersymmetric partners of the $W^{+}$ and $W^{-}$ gauge fields, see Eq.~(\ref{w}).

The $\Psi$'s decay into (s)quarks, the Higgs and Higgsino particles, and the $L_3,~e_1,~e_2$ (s)leptons. The relevant interaction terms for these decays are
\begin{eqnarray} \label{fermint}
&{\cal L}& \supset g_W ({\tilde Q}^{*}_{u,i} {\bar \Psi}_1 P_L \Psi_{d,i} + {\tilde L}^{1 *}_{3} {\bar \Psi}_1 P_L \Psi^2_{l,3} + {\rm h.c.}) \, \nonumber \\
& + & g_W ({H}^{1 *}_{u} {\bar \Psi}_1 P_L \Psi^2_{H_u} + {H}^{1*}_{d} {\bar \Psi}_1 P_L \Psi^2_{H_d} + {\rm h.c.}) \, \nonumber \\
& + & g_W ({\tilde Q}^{*}_{d,i} {\bar \Psi}_2 P_L \Psi_{u,i} + {\tilde L}^{2 *}_{3} {\bar \Psi}_2 P_L \Psi^1_{l,3} + {\rm h.c.}) \, \nonumber \\
& + & g_W ({H}^{2 *}_{u} {\bar \Psi}_2 P_L \Psi^1_{H_u} + {H}^{2*}_{d} {\bar \Psi}_2 P_L \Psi^1_{H_d} + {\rm h.c.}) \, \nonumber \\
& + & {g_W \over \sqrt{2}} ({\tilde Q}^{*}_{u,i} {\bar \Psi}_3 P_L \Psi_{u,i} - {\tilde Q}^{*}_{d,i} {\bar \Psi}_3 P_L \Psi_{d,i} + {\rm h.c.}) \, \nonumber \\
& + & {g_W \over \sqrt{2}} ({\tilde L}^{1 *}_{3} {\bar \Psi}_3 P_L \Psi^1_{l,3} - {\tilde L}^{2 *}_{3} {\bar \Psi}_3 P_L \Psi^2_{l,3} + {\rm h.c.}) \, \nonumber \\
& + & {g_W \over \sqrt{2}} ({{H}^{1*}_{u} {\bar \Psi}_3 P_L \Psi^1_{H_u}} - {H}^{2*}_{u} {\bar \Psi}_3 P_L \Psi^2_{H_u} + {\rm h.c.}) \, \nonumber \\
& + & {g_W \over \sqrt{2}} ({{H}^{1*}_{d} {\bar \Psi}_3 P_L \Psi^1_{H_d}} - {H}^{2*}_{d} {\bar \Psi}_3 P_L \Psi^2_{H_d} + {\rm h.c.}) \, \nonumber \\
& + & {\sqrt{2} g_Y \over 6} ({\tilde Q}^{*}_{u,i} {\bar \Psi}_4 P_L \Psi_{u,i} + {\tilde Q}^{*}_{d,i} {\bar \Psi}_4 P_L \Psi_{d,i} + {\rm h.c.})\, \nonumber \\
& + & {\sqrt{2} g_Y \over 2} ({\tilde L}^{1 *}_{3} {\bar \Psi}_4 P_L \Psi^1_{l,3} + {\tilde L}^{2 *}_{3} {\bar \Psi}_4 P_L \Psi^2_{l,3} + {\rm h.c.})\, \nonumber \\
& + & {\sqrt{2} g_Y \over 2} ({{H}^{1*}_{u} {\bar \Psi}_4 P_L \Psi^1_{H_u}} + {H}^{2*}_{u} {\bar \Psi}_4 P_L \Psi^2_{H_u} + {\rm h.c.}) \, \nonumber \\
& + & {\sqrt{2} g_Y \over 2} (- {{H}^{1*}_{d} {\bar \Psi}_4 P_L \Psi^1_{H_d}} - {H}^{2*}_{d} {\bar \Psi}_4 P_L \Psi^2_{H_d} + {\rm h.c.}) \, \nonumber \\
& + & {\sqrt{2} g_Y \over 6} (-4{\tilde u}_{i} {\bar \Psi}_{4} P_R \Psi_{u,i} + 2 {\tilde d}_{i} {\bar \Psi}_{4} P_R \Psi_{d,i} + {\rm h.c.}) \, \nonumber \\
& + & {\sqrt{2} g_Y \over 2} (2 {\tilde e}_{1} {\bar \Psi}_{4} P_R \Psi^2_{l,1} + 2 {\tilde e}_{2} {\bar \Psi}_{4} P_R \Psi^2_{l,2} + {\rm h.c.}) \, .
\end{eqnarray}
The rate for the decay of $\Psi_{1,2,3}$ to a massless scalar and its fermionic partner is $g^2_W M/32 \pi$, while that of $\Psi_4$ is $g^2_Y y^2 M/32 \pi$. Here $M$ denotes the mass of $\Psi$ and $y$ is the hypercharge of the scalar (and its fermionic partner).


\end{document}